\documentclass[twocolumn,aps,prb,superscriptaddress]{revtex4-2}

\usepackage{graphicx}
\usepackage{amssymb}

\begin{document}

\preprint{preprint}

\title{I. Magnetic, thermal and transport properties of YbCu$_{5-x}$Zn$_x$ alloys}



\author{I. Čurlik}
\affiliation{Faculty of Sciences, University of Prešov, 17. novembra 1, SK - 080 78 Prešov, Slovakia}

\author{F. Akbar}
\affiliation {UMR7140–CMC, Univ. Strasbourg, 
4 B. Pascal, CS 90032, 67081 Strasbourg, France}

\author{S. Gabani}
\affiliation{Institute of Experimental Physics, SAS, Watsonova 47, SK - 040 01 Košice, Slovakia}

\author{M. Giovannini}
\affiliation{Department of Chemistry, University of Genova, Via Dodecaneso 31, Genova, Italy}

\author{J.G. Sereni}
\affiliation{Low Temperature Division, CAB-CNEA, CONICET, IB-UNCuyo, 8400 Bariloche, Argentina}

\email[]{jsereni@yahoo.com}

\date{\today}

\begin{abstract}

Within the family of cubic YbCu$_4$X compounds ($X$ = Ni, Au and Zn), we have investigated the YbCu$_{5-x}$Zn$_x$ ($1\geq x \geq 0.7$) alloys by means of structural, magnetic, thermal and 
transport measurements.  
In the $\tau 1-$ YbCu$_{5-x}$Zn$_x$ (cubic AuBe$_5$ type, $0.7 \leq x \leq 1.5$) structural phase, Yb ion is in its Yb$^{3+}$ magnetic configuration. However, by increasing Zn content  the unit cell 
grows faster than a reference computed as a Cu by Zn atoms 
substitution, which indicates a shift of Yb ions towards the larger Yb$^{2+}$ configuration.
The magnetic behavior confirms such tendency with a clear decrease of the saturation magnetization and effective moment between $x=0.7$ and $x=1$. 
The specific heat at  low temperature shows a logarithmic dependence characteristic for a non-fermi-liquid behavior. 
The characteristic energies of  all studied parameters, including magneto resistivity, show notably low values as an indication that these alloys are close to a quantum critical point, which is approached 
from the non-magnetic side as the Zn content decreases.

\end{abstract}


\maketitle


\section{Introduction}

The family of Yb-based ternary intermetallic compounds YbCu$_4X$, with $X$= Ag, Au, Cd, Mg, Tl and Zn, is known since more than two decades \cite{Serrao}. It exhibits a cubic crystalline structure derivative of AuBe$_5$, sp. gr. F43m one \cite{Cubic}. This fcc lattice can be viewed as a network of edge-sharing tetrahedra with Yb magnetic ions 
located at the vertices, being a 3D analog of a triangular lattice \cite{ActPol}. The composition phase diagram shows the formation of solid solutions \cite{XtlChem} 
among which  $X$ = Au, Ag were intensively  studied 
\cite{YbCu5-xAux, YbCuAg, YbCuIn}. Particularly, YbCu$_{5-x}$Au$_x$ alloys have shown very interesting properties because of the lack of magnetic order even at very low temperature ($T\leq 40$\,mK) \cite{Banda} and its proximity to a quantun critical point (QCP).

In YbCu$_{5-x}$Au$_x$ allows the magnetic behavior is driven by the change in the chemical potential produced by the substitution of larger Au by smaller Cu rather than by their equivalent 
electronic structure, the question arises about which is the effect of the alternative experimental parameter like the change of electronic concentration. The recently synthesized YbCu$_{5-x}$Zn$_x$ 
solid solution \cite{Akbar} allows to investigate such effect. 

In this work we present a detailed study of the physical properties of YbCu$_{5-x}$Zn$_x$ within the $0.7\leq x \leq 1$ range of concentration obtained from structural, magnetic, thermal and transport 
properties. 
\section{Experimental}

\subsection{Structural properties} 
\begin{figure}
\begin{center}
\includegraphics[width=18pc]{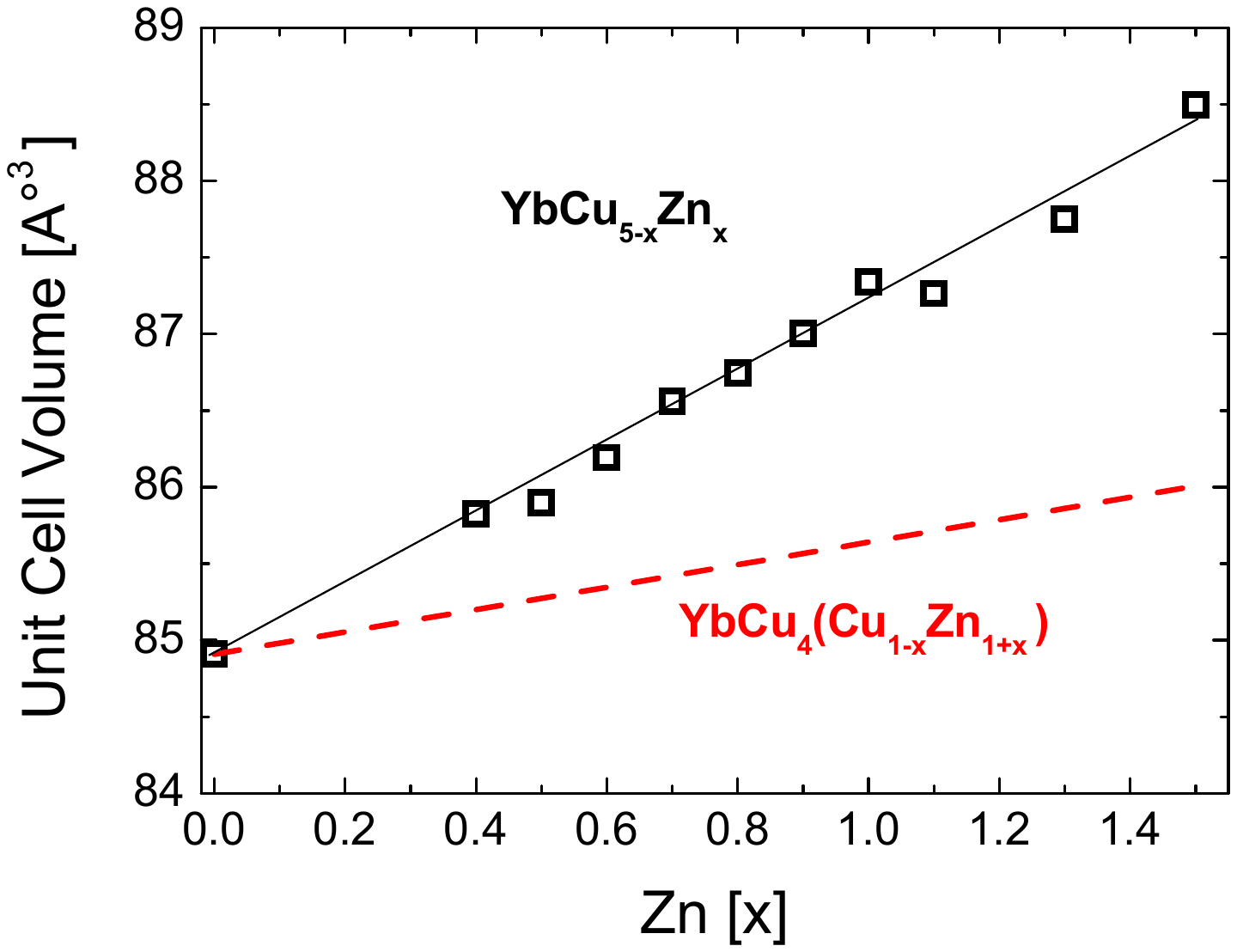}
\caption{(Color online) Comparison of the formula unit (f.u.) volume dependence with an hypotetical Vegard's law reference, where one (of five) Cu atoms is progressively substituted by one Zn. The 
Zn(x=0)origin is taken from \cite{YbCu5}. Notice that there are four f.u. in each cell volume \cite{Akbar}. \label{F1}}
\end{center}
\end{figure}
In most of these intermetallic compounds Yb atoms are in their Yb$^{3+}$ electronic configuration, they may enter into the range of valence instability when the energy of the $4f$ level ($E_{4f}$)
is close to the conduction band Fermi energy ($E_F$) through the $4f-band$ hybridization \cite{Wohll,Lang}. In that case, due to the larger size of Yb$^{2+}$ ions a supplementary expansion 
to the expected Vegard's law occurs. Due to the lack of a reference for such a Vegard's law volume evolution (e.g. a LuCu$_{5-x}$Zn$_x$ compound), we compare in Fig.~\ref{F1} the measured unit 
cell volume $V(x)$ increase as a function of Zn content with an `ad hoc' Vegard's reference. There, one (on five) Cu atoms are progressively 
substituted by one $Zn$. For this purpose the metallic radio of pure Cu and Zn elements is used and the Zn($x=0$) origin is extracted from \cite{YbCu5}.

One can see that $V(x)$ of YbCu$_{5-x}$Zn$_x$ increases quite faster than the expected from the 
size difference between Cu and Zn atoms, that is represented by the proposed Vergard's low for the 
pseudo YbCu$_4$(Cu$_{1-x}$Zn$_x$)  alloy. This evidence reveals the effect of mean sizes  
increase of Yb atoms towards the larger Yb$^{2+}$ configuration due to valance instability \cite{Valinst} or valence fluctuations \cite{Wohll,Lang}.

\begin{figure}
\begin{center}
\includegraphics[width=20pc]{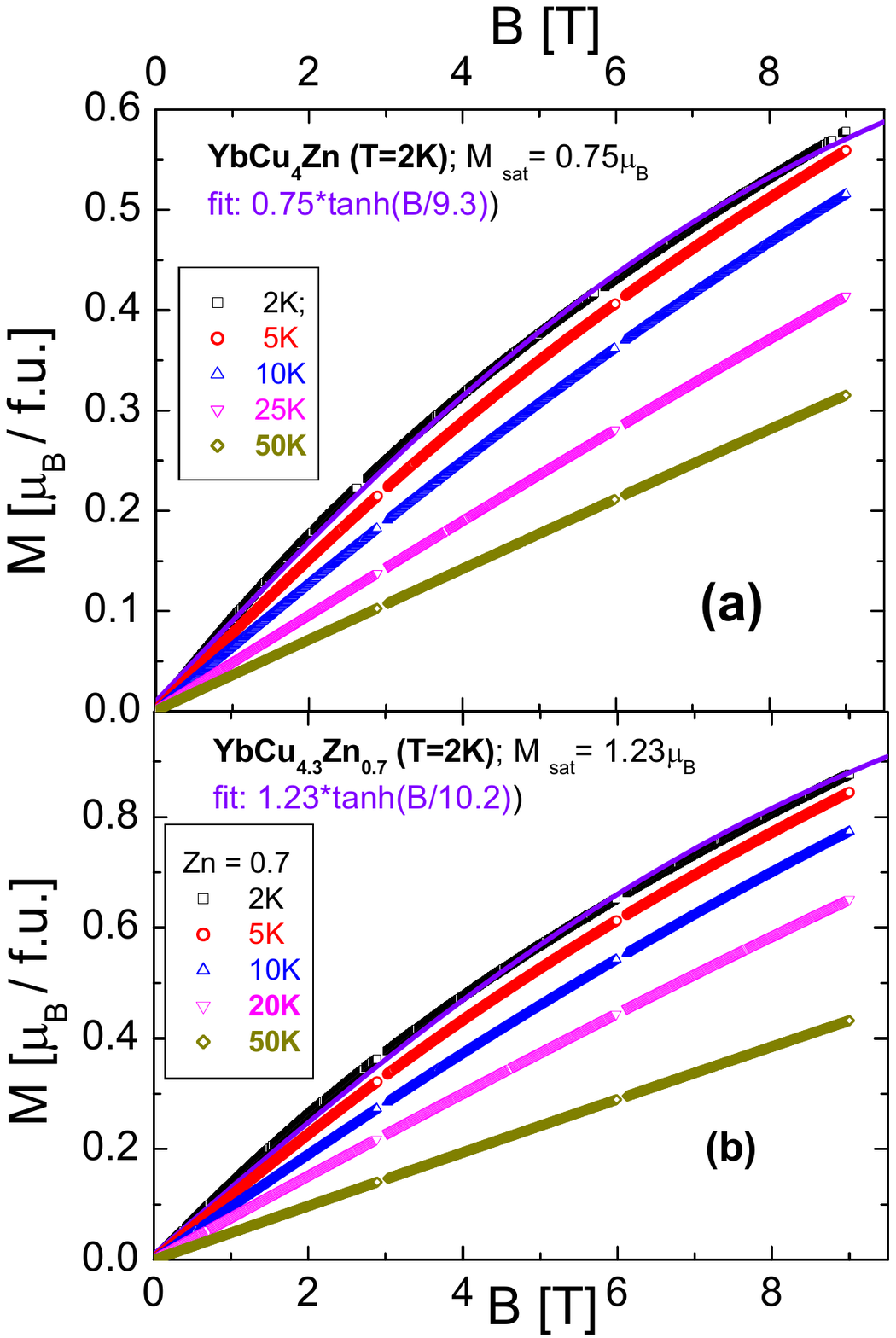}
\caption{(Color online) field dependent magnetization $M(B)$ for (a) YbCu$_4$Zn and (b) YbCu$_{4.3}$Zn$_{0.7}$, up to 9T in the range of 2\,K to 50\,K. Continuous curves in the 2\,K isotherms are the fits to extract the saturation magnetization. \label{F2}}
\end{center}
\end{figure}


\subsection{Magnetic properties}

\begin{figure}
\begin{center}
\includegraphics[width=20pc]{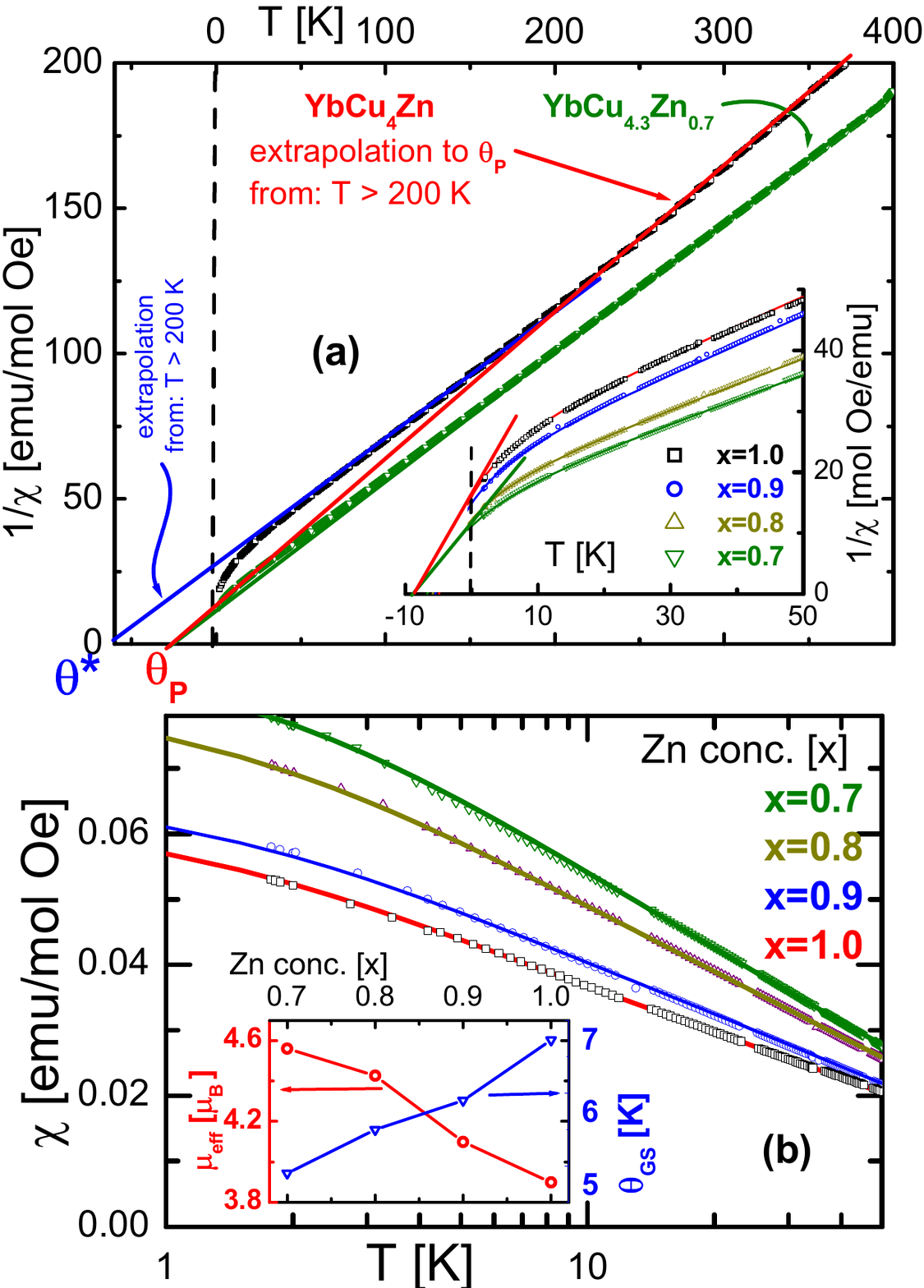}
\caption{(Color online) a) High temperature inverse susceptibility of YbCu$_{5-x}$Zn$_x$ (x=1 and x=0.7) alloys, up to room temperature, measured at $B=1$\,T (after \cite{Akbar}) where different extrapolations to $\theta_P^{ht}$ are discussed. 
Inset: detail of the inverse susceptibility at low temperature for all studied concentrations. 
b) Low temperature susceptibility in a semi-logarithmic representation, showing respective fittings. Inset: main extracted parameters: $\mu_{eff}$ (left blue axis) and $\theta_P^{lt}$  (right red axis). The left red arrow indicates the theoretic $\mu_{eff}(x)=4.54\,\mu_B$ value. \label{F3}}
\end{center}
\end{figure}

\subsubsection{Magnetization}

In Fig.~\ref{F2} we show the field dependent magnetization $M(B)$ for YbCu$_4$Zn (a) and YbCu$_{4.3}$Zn$_{0.7}$ (b), between 2 and 50\,K and up to $B=9$\,T.
The respective isotherms at 2\,K are properly fitted with a Brillouin function $B_{1/2}(y)$ (continuous curves), where  
$y = g_{eff} \mu_B J B/ k_B T$ \cite{Blundell}  for the present doublet GS. These functions are normalized to a saturation value of $M_{sat}= 0.75\,\mu_B$ for $x=1$ and 1.23$\mu_B$ for $x=0.7$. 

We can compare the highest saturation value (observed for the $x=0.7$ alloy) with those of the two possible GS doublets: $\Gamma_6$ and $\Gamma_/$, with respective $M_{sat}(\Gamma_6) 
=1.33\mu_B$ and $M_{sat} (\Gamma_7) =1.72\,\mu_B$.  It is evident that  $\Gamma_6$ is the most likely GS. The significant decrease of $M_{sat}(x)$ indicate a weakening of the magnetic character of these alloys by increasing the Zn content because this system is located on the non-
magnetic side of a quantum critical point (QCP), with the tendency to a fermi liquid behavior.  

\subsubsection{Magnetic susceptibility}

The magnetic susceptibility $\chi(T)$ of the YbCu$_{5-x}$Zn$_x$ alloys was measured up to room temperature with $B=1$\,T and the results are presented in 
Fig.~\ref{F3}a (for x=1 and x=0.7) as the inverse $1/\chi(T)$, after \cite{Akbar}. In a detailed analysis  of the crystal electric field CEF effect, one can see that it does not strictly follow a straight 
line: $1/\chi(T)= (T+\theta_P) /C_C$ as expected for the pure Curie-Weiss model. With $\theta_P$ being the paramagnetic 
temperature and $C_C$ the Curie constant $\propto 
\mu_{eff}^2$. This is because the GS degeneracy is broken by the lower symmetry of the CEF. 
Fig.~\ref{F3}a shows a moderate upturn of measured $1/\chi(T)$for YbCu$_4$Zn, which is related to 
the progressive depopulation of the excited CEF levels by cooling, see red straight line which extrapolates to $\theta_P$. This 
feature is less pronounced for YbCu$_{4.3}$Zn$_{0.7}$ because of the atomic disorder at the $4c$ Wyckoff site \cite{Akbar} introduced by the random occupation of Cu and Zn by alloying.  
Therefore, for a precise evaluation of $\theta_P$ the high temperature extrapolation has to be done from $T\geq 200$\,K in this case, because a lower temperature e.g. $T\leq 200$\,K (see blue 
line in  Fig.~\ref{F3}a) extrapolation would result in an artificial over-evaluation of $\theta^* \approx 60$\,K. The value $\theta_P= 26\pm 2\,K$ coincides for all Zn concentrations. 
From the Kondo temperature evaluation: $T_K = \theta_P/\sqrt 2$ \cite{Katanin}, one extracts for high temperature $T_K = 15$\,K that starts to affect the contribution of the excited CEF 
levels \cite{Desgranges}. 

\begin{figure}
\begin{center}
\includegraphics[width=20pc]{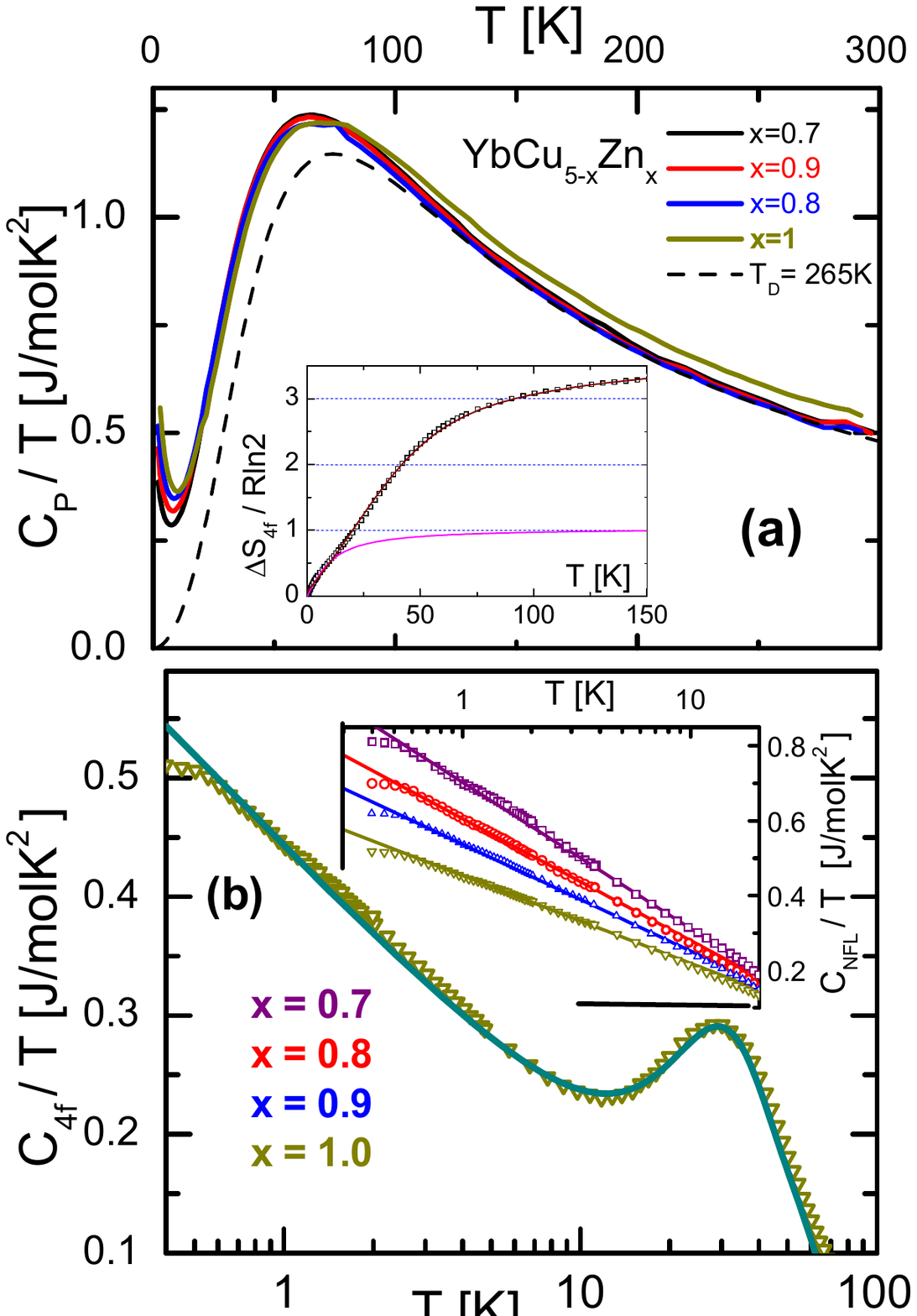}
\caption{(Color online) a) High temperature specific heat of YbCu$_{5-x}Zn_x$ alloys up to room temperature in a $C_P/T$ representation, compared with a Debye function of $T_D=265$\,K (dashed 
curve). Inset: Entropy of the $4f$ electrons after phonon subtraction. Continuous curves indicate the  increase to the entropy of respective CEF levels: excited doublet (magnet) and $\Gamma_8$ quartet 
(dark yelow) . 
b) Specific heat contribution divided temperature of the $4f$ electrons in a semi-logarithmic representation indicating the NFL behavior of the GS doublet for the YbCu$_4$Zn$_1$ compound from 
$T\geq 0.4$\,K up to 10\,K. Continuous curve: fit between 0.4 and 70\,K (see the text) including the CEF contribution.
Inset: Comparison of the $C_{NFL}/T$ dependencies of the YbCu$_{5-x}$Zn$_x$ alloys. \label{F4}}
\end{center}
\end{figure}


To analyze the GS properties of these compounds, in the inset of Fig.~\ref{F3}  we include more details on the $T\leq 50$\,K data for all the Zn concentrations. In order to better compare their  characteristic parameters, we applied in the inset of Fig.~\ref{F3}a 
an heuristic equation:
\begin{center}
$1/\chi(T, x)=[T+\theta_{GS}] / C_C\, e^{-\Delta/[T+\theta_P]}$
\end{center}
where $\theta_{GS}(x)$ characterizes the GS paramagnetic temperature, $C_C(x)$ are the respective Curie constants and $\Delta(x)$ proportional to the CEF splitting. In  Fig.~\ref{F3}b the $\chi(T)$ is 
also presented to show the quality of the fit at low temperature. 

Another check for these results can be done by comparing the $\mu_{eff}(GS)$ extracted from $1/\chi$ at $T\approx 2$\,K with the magnetic saturation at that temperature because $\mu_{eff}(GS)/
M_{sat}(2\,K) =2.13\mu_B/1.23 \mu_B\approx \sqrt 3$, as expected for a doublet ($S=1/2$) ground state. 

The effective magnetic moments $\mu_{eff}(x)$ depend on Zn concentration as displayed in the inset of Fig.~\ref{F3}b and range between 
 $\mu_{eff}(0.7) = 4.57\,\mu_{B}$ and  $\mu_{eff}(1.0) = 3.95\,\mu_{B}$.  Notice that the theoretic effective moment for $
\vec{J}=7/2, \vec{S}=1/2, \vec{L}=3$ of Yb atoms is $\mu_{eff} 
= g_J[J(J+1)]^{1/2} = 4.54\,\mu_B$, see dashed (red) line in the inset of Fig.~\ref{F3}b, with a  Landé factor $g_J=7/6$.
 
The inset also contains the values of $\theta_{GS}$ obtained by extrapolating the  
$1/\chi(T\approx 2K)$ curves to zero. These parameter ranges between $\theta_{GS}(0.7)=4.7$\,K and $\theta_{GS}(1.0) =7$\,K. The CEF splitting slightly change from $\Delta(0.7)= 42$\,K to $\Delta(1.0)= 38$\,K, like $\theta_P$ between 23 and 26\,K.


\subsection{Specific heat} 
\subsubsection{High temperature specific heat}

The specific heat of YbCu$_{5-x}Zn_x$ alloys are shown in Fig.~\ref{F4}a also up to room temperature, in a $C_P/T$ representation and compared with a pure phonon-Debye function 
($C_D$) with $T_D=265$\,K (dashed curve). The $T_D$ value was determined by the temperature of the maximum of $C_D/T = 0.28 T_D$ \cite{SpecHeat}, that is in good agreement with the value 
obtained for LuCu$_{5-x}$Ag$_x$ \cite{LuCuAg}. As expected, no significant variation in the 
phonon spectrum is observed as a function of Zn concentration because of its similar weight as Cu. 

From the difference: $C_{4f}/T = C_P/T - C_D/T$ the $4f$ electronic contribution of the crystal electric field CEF is extracted. This procedure is checked by computing the $4f$ entropy $\Delta_{4f}/
Rln2$ 
accumulated up to 150\,K, see the inset in Fig.~\ref{F4}a, where the continuous curves represent the entropy variation related to the excited doublet (magenta) and the total entropy including the $
\Gamma_8$ quartet. The excess of entropy at high temperature is originated in the band electrons and corresponds to a low Sommefeld coefficient $\gamma \approx 5$\,mJ/mol K$^2$, which for LuCu$_{5-x}$Ag$_x$ \cite{LuCuAg} was reported to be $\gamma \approx 9$\,mJ/mol K$^2$

\subsubsection{Intermediate and low temperature specific heat}

In Fig.~\ref{F4}b we present the $C_{4f}/T$ contribution to specific heat for YbCu$_4$Zn in a semi-logarithmic representation, covering more than two orders of magnitude in temperature. There, it is evident 
the dominant NFL behavior of the GS doublet between $T\geq 0.4$\,K up to 10\,K, followed by a the CEF excited levels contribution. The $10\leq T \leq 100$\,K anomaly cannot be properly fitted by a 
standard Schottky one 
because those levels are affected by Kondo interaction that induces a level broadening. Such effect is described using a Gaussian function centered on the CEF level energy and a dispersion factor. The 
used function to fit $C_{4f}/T$ is the following: 
\begin{equation}  
a+b\log(T_{q}/T)+\Sigma_{i=1}^2 c_i\,e^{[-(T-\Delta_i)^2/(2*T_{Ki}^2)]}
\end{equation}
This equation is composed by terms with different physical meaning. The constant value $a=0.15$\,J/molK$^2$ is originated in the electronic band contribution. At low temperature $C_{4f}/T =0.26 
\log(T_{q}/T)$ is characteristic of a NFL behavior for the GS doublet, with a quantum 
fluctuation energy of $T_{q} = 7.5$\,K. 

The third term in eq.(1) includes two Lorentzian functions that account for the excited CEF levels 
From this fit we extract the CEF levels splitting of $\Delta_1 =30\pm 2$\,K to the first excited doublet  and a width $T_{K1}=15 \pm 1$\,K, in good agreement with the $T\leq 30$\,K curvature of $1/
\chi(T)$ in Fig.~\ref{F3}. 
The values for the upper level are $\Delta_2 = 46\pm 3$\,K and a width $T_{K2}=40 \pm 2$\,K. The $c_i$ prefactor is related to the degeracy of each level, where the quarted doubles the first excited 
doublet as expected. 
The full fit is presented in Fig.~\ref{F4}b as a continuous (green) function in the range between 0.4 and 70\,K.  

The CEF levels splitting is relatively low in comparison to some other Yb-based compounds. Nevertheless, in this case the cubic crystalline structure and the similar 
electronic structure of Cu and Zn provides a quite symmetric environment to Yb ions. Both features provide highly symmetric environment to Yb ions.  
These values extracted from $C_{4f}(T)$ measurements can be compared with those from $\chi(T)$ results, but taking into account that the former are more precise because they are related to the 
derivative of the internal energy of the system (e.g. $C_{4f}=\partial U_{4f}/\partial T$). 

\begin{figure}
\begin{center}
\includegraphics[width=20pc]{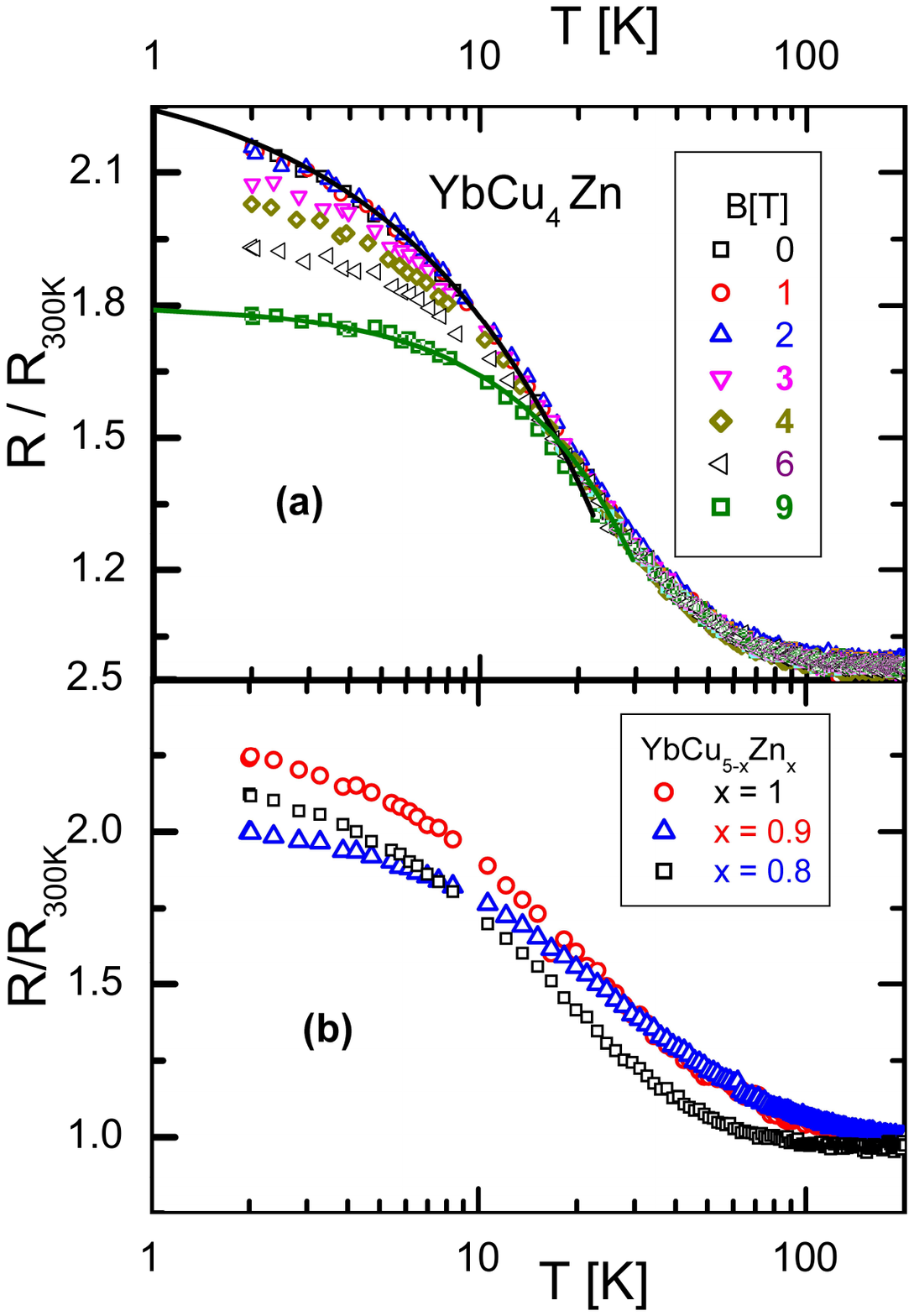}
\caption{(Color online) Temperature dependence of the electrical resistivity of YbCu$_4$Zn,  normalized to room temperature in a semi-logarithmic representation. a) YbCu$_4$Zn compound at different applied fields. Continuous curves are fits to 
describe the curvature evolution at low temperature, see the text. b) Substitution of Cu by Zn in YbCu$_{5-x}$Zn$_x$\label{F5}}
\end{center}
\end{figure}


Concerning the GS behavior, in the inset of Fig.~\ref{F4}b we compare the $C_{NFL}/T$ dependencies of the YbCu$_{5-x}$Zn$_x$ alloys for $1\geq x \geq 0.7$. These measurements were 
preformed in two different calorimeters, those at $T<3$\,K in a He3 insert of PPMS (Quantum Design) and those at $T>2$\,K in an He4 pumped system. Therefore respective results were matched 
between 2 and 3\,K adding a small correction at the $T>3$\,K data.
 Notice that for the $C_{NFL}/T = b\log(T_{Q}/T)$ function, the 
characteristic energy of quantum fluctuations $T_{Q} = 9$\,K do not 
change with concentration. Nevertheless, the intensity of the $C_{NFL}/T$ contribution increases by reducing Zn content, i.e. increasing the magnetic character of the alloys that suggests an increasing 
proximity to a QCP. 


\subsection{Electrical resistivity and Magnetoresistivity}

\begin{figure}
\begin{center}
\includegraphics[width=20pc]{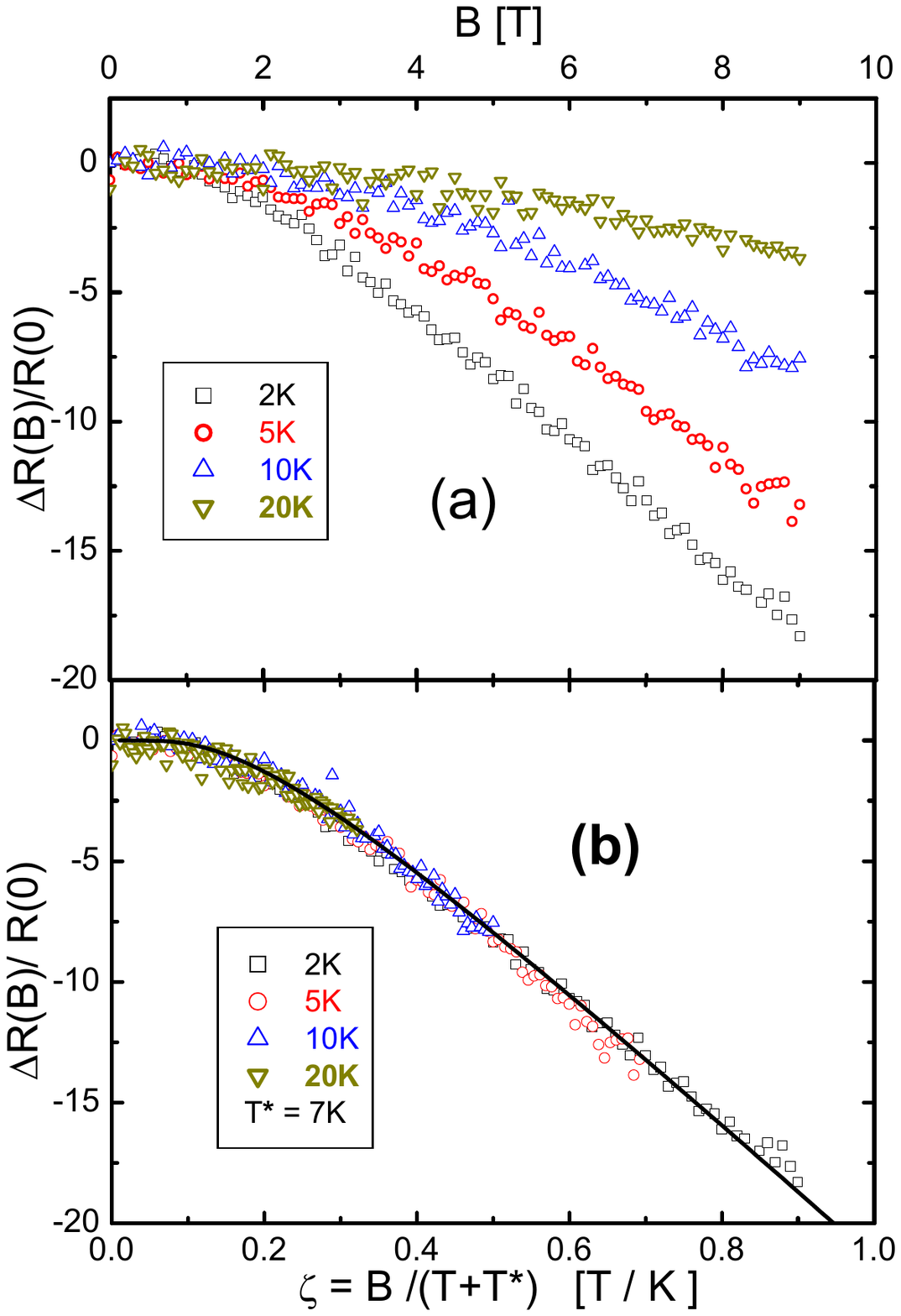}
\caption{(Color online) a) Magnetoresistivity measurements on YbCu$_4$Zn at different temperatures. b) Scaling procedure with a characteristic temperature scale $T^*=7$\,K using curves at $T=2, 5, 10$ and 20\,K (see text). Continuous curve: fit with $r(B,T)= -29\,\zeta\,\,  \exp{(-0.3/\zeta)}$.\label{F6}}
\end{center}
\end{figure}

The electrical resistivity of some selected Zn alloys was measured between 2\,K and room temperature. In the stoichiometric compound it was measured in magnetic field up to $B=9$\,T and presented in a semi-
logarithmic representation with the resistivity values normalized to room temperature $R/R_{300k}$, see  Fig.~\ref{F5}a. The general 
behavior resembles that of a single impurity system. There one can appreciate how the electronic scattering tends to saturation at low temperature, but without any evidence of coherency at $T\to 0$.

The resistivity for different Zn concentrations are collected in Fig.~\ref{F5}b. Despite of the values dispersion between different alloys, one can see a similar qualitative temperature dependence.   


The reduced magnetoresistivity $r(B)=\Delta R(B)/R(0)$ of YbCu$_4$Zn in magnetic field  up to 9\,T is presented in Fig.~\ref{F6}a, showing a negative tendency as a function of field. From the curves at $T= 2, 5, 10$\,K a scaling 
procedure proposed by Schlottmann \cite{Schlott} can be performed. 
In Fig.~\ref{F6}b we show the unified curve of $r(T, B)$ dependence on $ \zeta =B/(T+T^*)$ from where the characteristic temperature $T^*= 7$\,K is extracted. 
The scaling function is described as: $r(B,T)= -2\,\zeta\,\, exp{(-0.3/\zeta)}$.

\section{Conclusions}

From these four experimental parameters: magnetic susceptibility, specific heat, entropy and  magnetoresistivity, it is evident that the doublet GS of YbCu$_4$Zn has a significantly low 
characteristic energy scale, between 7 and 8.7\,K. This may occur because of the lack of long range collective interactions, topologically prevented by magnetic frustration

As a function of Zn concentration, we have determined that: i) the lattice expansion exceeds the expected from a sort of Vegard's law reference, that corresponds to a shift towards the larger size 
Yb$^{2+}$ electronic configuration. ii) the effective moment decreases from x=0.7 towards 1 indicating the weakening of magnetism. iii) Coincidentally, the paramagnetic 
temperature slightly increases that may indicate an increase of the Kondo interaction. iv) also the low temperature specific heat intensity decreases suggesting a departure from an eventual QCP as the 
electronic concentration increases driving the Yb ions towards the less magnetic configuration. v) Furthermore, the entropy of the GS doublet reaches the value of $1/2R\ln(2)$ at 8.7\,K which is also 
considered a measure of the Kondo energy. Altogether one can conclude that YbCu$_{5-x}$Zn$_x$ is an appropriate system for the study of the environment of a QCP.

\section*{Acknowledgements}

We acknowledge the support from APVV-23-0226 and VEGA 2/0034/24 projects.

\end{document}